\documentclass[english]{ieej-e}
\usepackage{graphicx}
\usepackage[fleqn]{amsmath}
\usepackage[varg]{txfonts}
\FIELD{B}
\YEAR{2024}
\title{Rethinking Static Line Rating for Economic and Efficient \\ Power 
Operation in South Korea}

\authorlist{%
\authorentry{Junseon PARK}{n}{*}
\authorentry{Junhyun LEE}{n}{*}
\authorentry[hyeongon@pknu.ac.kr]{Hyeongon PARK}{n}{*} 
}
\affiliate[]
{School of Systems Management and Safety Engineering, College of Engineering,
Pukyong National university 45 Yongso-ro, Namgu, Busan 48513, Republic of Korea}

\received{2024}{11}{29}
\begin{document}
\begin{abstract}
In South Korea, power grid is currently operated based on the static line rating (SLR) method, where the transmission line capacity
 is determined based on extreme weather conditions. However, with global warming, there is a concern that the temperatures during summer
  may exceed the SLR criteria, posing safety risks. On the other hand, the conservative estimates used for winter conditions limit 
  the utilization of renewable energy. Proposals to install new lines face significant financial and environmental hurdles, 
  complicating efforts to adapt to these changing conditions. Dynamic Line Rating (DLR) offers a real-time solution 
  but requires extensive weather monitoring and complex integration. This paper proposes a novel method that improves on SLR by analyzing
   historical data to refine line rating criteria on a monthly, seasonal, and semi-annual basis. Through simulations, 
   we show our approach significantly enhances cost-effectiveness and reliability of the power system, 
   achieving efficiencies close to DLR with existing infrastructure. This method offers a practical alternative
    to overcome the limitations of SLR and the implementation challenges of DLR.

\end{abstract}
\maketitle

\section{Introduction}
In South Korea, the operation of transmission lines is regulated by the Static Line Rating (SLR) system, which dictates a constant current capacity
 that remains unchanged over time.
  This is based on criteria set by the KEPCO\cite{kepco_guidelines} - a maximum temperature of 40°C and a wind speed of 0.5 m/s - 
 to determine the maximum amount of current a transmission line can safely carry under the most severe weather conditions throughout the year.
  However, in recent years, as global warming has accelerated, there have been instances of extreme weather conditions exceeding these criteria 
  during the summer months when temperatures are relatively high\cite{hot_summer2018}. On the other hand, during the winter months, despite the temperatures 
  being significantly lower compared to 40°C, the criteria are set too conservatively, resulting in inefficient utilization of transmission network
   resources. To summarize, extreme weather conditions that exceed the criteria in summer have the potential to cause safety issues, 
   while conservative criteria in winter limit the efficient use of network resources.\\
Several prior studies have been conducted to address this issue. Building additional lines has been proposed as a way to increase transmission 
capacity, but this approach is difficult to implement due to financial, environmental, and social issues\cite{Distance,Conflicts}. As an alternative,
 much research has been done on Dynamic Line Rating (DLR), which regulates the amount of current that can flow through a transmission line 
 in response to changing weather conditions.
However, effective implementation of DLR presents technical challenges including the development 
 of real-time weather condition monitoring systems and the installation of weather sensors. In addition, the introduction of such systems 
 and equipment can entail financial costs. As a compromise, there's the option of selectively applying DLR to particular lines where
  it's especially effective. However, line congestion varies depending on system operation conditions, and predicting it in advance is computationally 
  expensive\cite{expensive}.
The authors in\cite{wind_power} shows that DLR can utilize power network resources more efficiently than SLR by analyzing the overall benefits as a function 
of the percentage of curtailed energy and the number of power plants. The study highlights the positive impact of DLR deployment on the operation
 of the power system. However, the study does not take into account the case of worse than usual weather conditions set by SLR, and 
 the operating cost analysis focuses only on curtailed energy, which is not realistic.\\
This study analyzes the limitations of the conventional SLR operation method from various aspects and proposes a new SLR operation method 
to improve the conventional SLR operation method. This study applies the concept of optimal power flow (OPF) to analyze the operating cost of
 the power system more realistically, considering the power generation cost of each power plant and the hourly varying power demand.
  Furthermore, we propose a novel approach to updating weather criteria on a monthly, seasonal, and six monthly basis, 
  which aims to simultaneously increase the safety and efficiency of the power system by being more conservative in high temperatures and relaxed 
  in low temperatures. Our analysis not only demonstrates the feasibility but also highlights the benefits of adopting the new SLR operation method.\\
 This paper is organized as follows. Chapter 2 briefly introduces the methodology for calculating the allowable current of transmission lines. 
 Chapter 3 systematically analyzes the problems of the conventional SLR operation focusing on safety and efficiency. 
 Chapter 4 conducts simulations to evaluate the operating costs of a new approach that applies the OPF concept, 
 aiming to optimize the power grid's performance by considering updates to weather criteria on a monthly, seasonal, and six-monthly basis.
  Finally, Chapter 5 discusses the contributions of this paper and discusses the limitations of this research.

\section{How to Calculate Transmission Line Allowable Current}
The method for determining the allowable current for transmission lines is based on the IEEE 738 document \cite{IEEE}.
This standard outlines methods for calculating the allowable current based on weather data, including wind speed and temperature.
The specific calculation is determined by the following formula:
\begin{eqnarray}
I = \sqrt{\frac{q_c + q_r - q_s}{R(T_{\text{avg}})}}    
\end{eqnarray}
\begin{eqnarray}
q_{c1} = K_{angle}[1.01 + 1.35N_{Re}^{0.52}]k_{f}(T_{s} - T_{a})  
\end{eqnarray}
\begin{eqnarray}
q_{c2} = 0.754K_{angle}N_{Re}^{0.6}k_{f}(T_{s} - T_{a})    
\end{eqnarray}
\begin{eqnarray}
q_{cn} = 3.645\rho_{f}^{0.5}D_{0}^{0.75}(T_{s} - T_{a})    
\end{eqnarray}
\begin{eqnarray}
q_{r} = 17.8D_{0}\epsilon[(\frac{T_{s} + 273}{100})^4 - (\frac{T_{a} + 273}{100})^4]    
\end{eqnarray}
\begin{eqnarray}
q_{s} = \alpha Q_{se}(\sin{\theta})A'
\end{eqnarray}
Table 1. describes the nomenclature used to obtain the allowable current.
\begin{table}[tb]
\begin{center}
\caption{Nomenclature for calculating allowable current}
\label{tab:symbols_descriptions}
\begin{tabular}{c|p{6.8cm}}
\hline
\textbf{Symbols} & \textbf{Description}\\
\hline
$I$ & Conductor current\\
\hline
$q_r$ & Radiated heat loss rate per unit length\\
\hline
$q_s$ & Heat gain rate from sun\\
\hline
$R(T_{avg})$ & AC resistance of conductor at temperature $T_{avg}$\\
\hline
$q_{c1}$ & Convection heat loss rate per unit length at low wind speed\\
\hline
$q_{c2}$ & Convection heat loss rate per unit length at high wind speed\\
\hline
$q_{cn}$ & Convection heat loss rate per unit length at zero wind speed\\
\hline
$K_{angle}$ & Wind direction factor\\
\hline
$N_{Re}$ & Reynolds number\\
\hline
$k_f$ & Thermal conductivity of air at temperature $T_{film}$\\
\hline
$T_s$ & Conductor surface temperature\\
\hline
$T_a$ & Ambient air temperature\\
\hline
$V_w$ & Wind Speed\\
\hline
$\rho_f$ & Density of air\\
\hline
$D_0$ & Outside diameter of conductor\\
\hline
$\epsilon$ & Emissivity\\
\hline
$\alpha$ & Solar absorptivity \\
\hline
$Q_{se}$ & Total solar and sky radiated heat intensity corrected for elevation \\
\hline
$\theta$ & Effective angle of incidence of the sun's rays \\
\hline
$A'$ & Projected area of conductor \\
\hline
\end{tabular}
\end{center}
\end{table}    
Equation (1) determines the allowable current, whose factors include heat dissipation by convection and radiation,
absorption of solar heat, and the alternating resistance of the conductor measured at an average temperature. 
The resistance value changes with temperature, and in this study, we use the resistance value according to KEPCO's standards \cite{kepco_guidelines}.
Equations (2)-(4) calculate the heat dissipation by convection. They represent the heat dissipation equations for low wind speed,
high wind speed, and no wind speed, respectively, and the largest value of $q_{c1}, q_{c2}, q_{cn}$  is adopted to calculate the allowable current.
Equations (5) and (6) calculate the heat dissipation by radiation and solar heat absorption, respectively. 
The parameter values used to calculate the allowable current in Korea and the corresponding results of the allowable currents 
are summarized in Table 2.
\begin{table}[tb]
\begin{center}
\caption{Parameter values used in South Korea}
\label{tab:parameter}
\begin{tabular}{c|p{6.8cm}}
\hline
\textbf{Symbols} & \textbf{Value used(KEPCO)}\\
\hline
$R(T_{avg})$ & 0.0804Ω/km\\
\hline
$K_{angle}$ & 1\\
\hline
$T_s$ & 90℃\\
\hline
$T_a$ & 40℃ at conventional SLR\\
\hline
$V_w$ & 0.5m/s at SLR\\
\hline
$D_0$ & 30.42mm\\
\hline
$\epsilon$ & 0.5\\
\hline
$\alpha$ & 0.5 \\
\hline
\end{tabular}
\end{center}
\end{table} 
It is emphasized that the air temperature($T_{a}$) is an important variable in the process of determining the allowable current 
and has a significant impact on the calculation of the allowable current.

\section{The Problems Of The Conventional SLR Operation}
\subsection{Safety issues of transmission lines due to accelerating global warming}
In South Korea, weather criteria for determining allowable current have been established at a temperature of 40°C 
and a wind speed of 0.5 m/s. These criteria were determined by analyzing the highest temperatures in 72 regions across 
the country from 1971 to 1999\cite{Distance}. However, as global warming accelerates, 
extreme weather conditions that exceed the criteria are emerging. This is particularly evident on August 1, 2018, 
when temperatures of 40.1°C and 41°C were recorded in Hongcheon, Gangwon-do, at 2pm and 4pm, respectively \cite{Data}. 
Figure 1. shows the result of calculating the allowable current of the ACSR 480C line type using hourly temperature data 
from Hongcheon, Gangwon-do, Korea in 2018, with the wind speed fixed at 0.5 m/s. 
\begin{figure}[tb]
\begin{center}
\includegraphics[width=\linewidth]{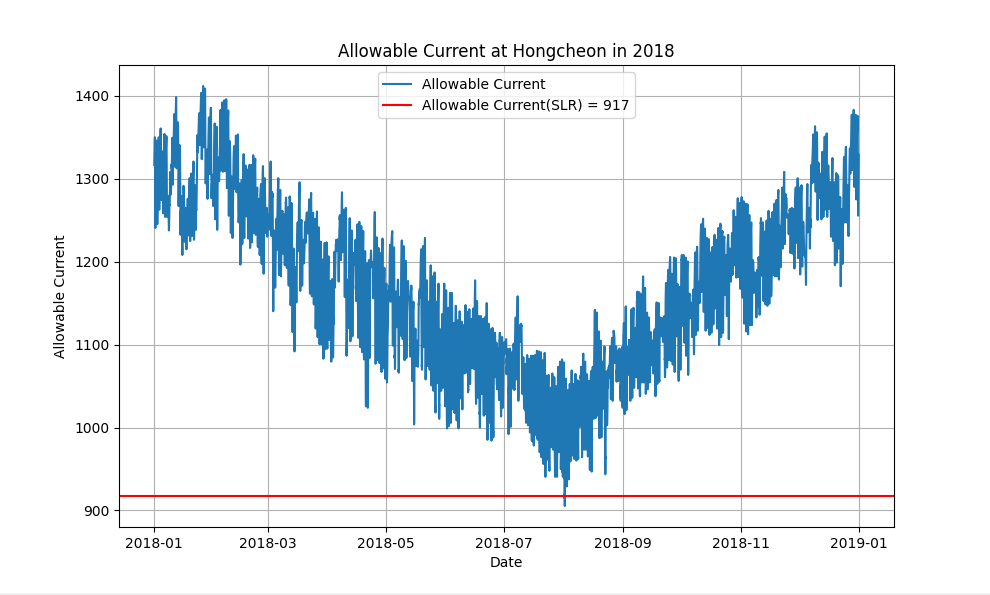}
\end{center}
\caption{Calculation of allowable current and current allowable current criteria for Hongcheon in 2018}
\label{fig:Figure1}
\end{figure}

In the Figure 1., the blue line shows the allowable current calculated by reflecting the real-time temperature in the Hongcheon area, 
and the red line shows the allowable current calculated based on the conventional operating weather conditions (40°C air temperature and wind speed of 0.5 m/s).
The graphs show that in extreme weather conditions with temperatures above 40°C, the allowable current can be lower than the 917A's reference calculated
under conventional weather criteria. This situation can lead to increased wear and tear or strain on 
the transmission lines and sag phenomena, which in turn can negatively impact safety. In addition, due to global warming, temperatures in Korea are rising.
As an example, Figure 2. is an analysis of the monthly maximum temperature in Busan for 50 years from 1974 to 2023.

\begin{figure}[tb]
\begin{center}
\includegraphics[width=\linewidth]{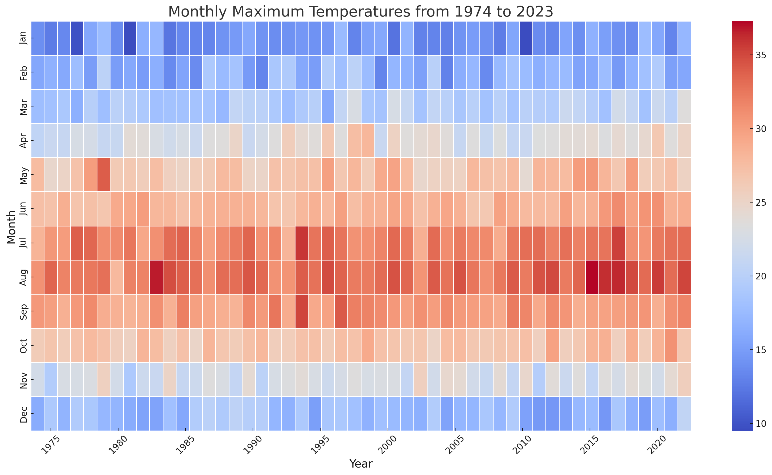}
\end{center}
\caption{Monthly maximum temperature in Busan over 50 years from 1974 to 2023}
\label{fig:Figure2}
\end{figure}

This is represented in the heatmap by the blue color scheme for lower temperatures and the red color scheme for higher temperatures. 
Of these, the change in temperature observed in August stands out. The heatmap shows that, over time, the representation of August temperatures has shifted towards 
progressively darker shades of red, indicating an increasing trend towards higher temperatures in this month. In another example, 
when analyzing the temperature change in each province in South Korea from 1974 to 2023, the largest temperature increase was in Miryang, Gyeongsangnam-do. 
Figure 3. shows the annual maximum temperature and its 3-, 5-, and 10-year moving averages from 1974 to 2023 for Miryang.

\begin{figure}[tb]
\begin{center}
\includegraphics[width=\linewidth]{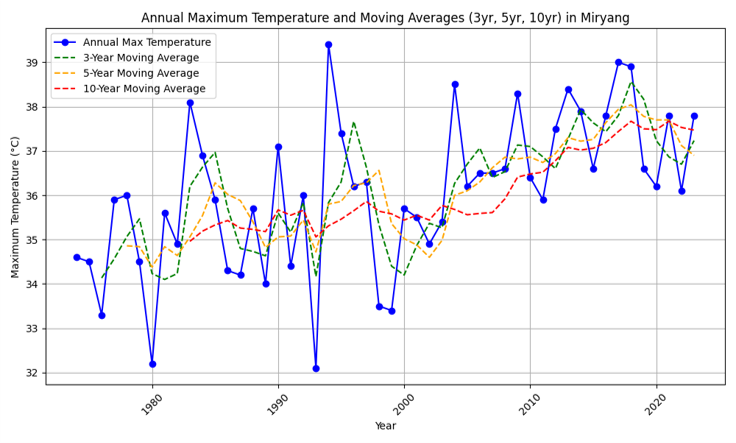}
\end{center}
\caption{Maximum temperature and 3, 5, and 10-year mov-ing averages for Miryang from 1974 to 2023}
\label{fig:Figure3}
\end{figure}

Figure 3. shows that maximum temperatures have been increasing over time, and this trend is expected to continue in the future. 
This trend of rising maximum temperatures suggests that, over time, the likelihood of surpassing the thresholds set by conventional weather criteria will increase. 
Consequently, relying on these traditional standards for operating the power system exposes it to a growing risk of inefficiency and potential failure.
 Adapting operational strategies to account for this changing climate is becoming increasingly important to ensure the reliability and safety of the power infrastructure.

\subsection{Inefficient operation of power systems at low temperatures}
While addressing safety concerns due to global warming is vital, it is equally imperative to consider the economic efficiency of 
power system operations. At relatively low temperatures, uti-lizing weather criteria set much higher than actual temperatures leads to 
the inefficient utilization of network resources. Figure 4. presents the calculated allowable current for an ACSR 480C line, with the wind speed set at 0.5 m/s.
 It uses the highest monthly temperatures, drawn from hourly data, for Hongcheon, Gangwon-do, from the year 2018.

 \begin{figure}[tb]
\begin{center}
\includegraphics[width=\linewidth]{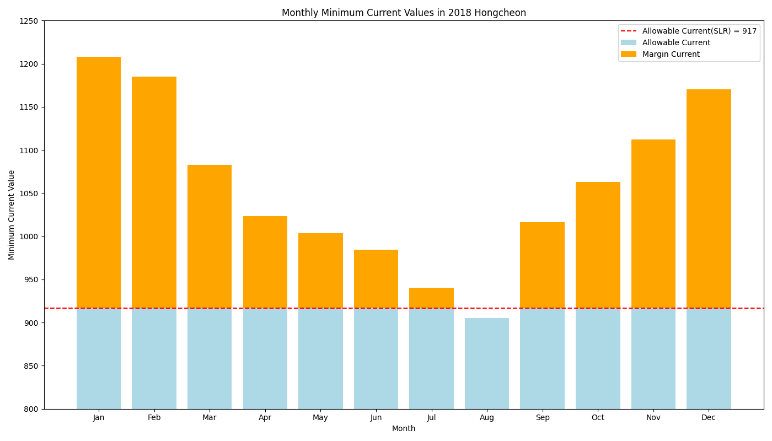}
\end{center}
\caption{Calculation of allowable current and current allowable current criteria for Honcheon in 2018}
\label{fig:Figure4}
\end{figure} 

According to the Figure 4., the blue section are the results of calculating the allowable current based on the criteria of conventional operating weather conditions,
 with 917A represented by the dotted line. The orange section shows the allowable current margin calculated by taking into account the maximum temperature and 
 wind speed of 0.5 m/s for each month. When the temperature is high, such as in August, it is more extreme than the weather criteria.
  On the other hand, during the colder months, such as January, February, and December, the weather criteria are set more conservatively than what is actually feasible. 
  This conservative setting prevents the full utilization of electric power network system. This case leads to the use of more expensive generators and the inability 
  to operate the power system economically.  

\section{Analyze Optimal Operating Costs}
\subsection{Simulation model formulation}
The goal of OPF analysis is to calculate the most efficient distribution of power considering power flow equations, trans-mission power limits,
 generator output constraints, and a balance between supply and demand, all while aiming to minimize the operating costs of the power system \cite{OPF,PowerFlowEq,SCUC}.
For a more realistic approach, the simulation considers the demand change of each bus every hour for a year (8760 hours), and the total operating cost for 24 hours (one day)
is minimized. Thus, the optimization was performed by repeating a problem with a scheduling horizon of 24 hours (one day) 365 times. 
The mathematical formulation is presented below, and the parameters used are listed in the table in Table 3..

\begin{table}[tb]
\begin{center}
\caption{Nomenclature for mathematical formulation}
\label{tab:symbols_descriptions2}
\begin{tabular}{c|p{6.8cm}}
\hline
\textbf{Symbols} & \textbf{Description}\\
\hline
$T$ & Total hours(24 hours)\\
\hline
$t$ & Time index\\
\hline
$G$ & Total number of generator\\
\hline
$g$ & Generator index\\
\hline
$P_{g}$ & Power generation from the generator g per 1 hour(MWh)\\
\hline
$C_{g}$ & Cost of generation from generator g (\$)\\
\hline
$M$ & Total number of buses\\
\hline
$i$ & Bus index\\
\hline
$j$ & Another bus index\\
\hline
$VOLL$ & Load curtailed cost\\
\hline
$P_{d}$ & Load power per 1 hour(MWh)\\
\hline
$P_{l}$ & Power flowing on the 1 line per 1 hours\\
\hline
$B_{ij}$ & Susceptance matrix elements from i-bus to j-bus\\
\hline
$\theta$ & Voltage angle\\
\hline
$L$ & Total number of line\\
\hline
\end{tabular}
\end{center}
\end{table}

\begin{eqnarray}
minimize, \sum_{t=1}^{T}(\sum_{g \in G}^{}C_{g} \cdot P_{g, t} + \sum_{i \in M}^{}VOLL \cdot P_{c, i, t})
\end{eqnarray}
subject to
\begin{eqnarray}
P_{g}^{min} \leq P_{g, t} \leq P_{g}^{max}\hspace{1cm}\forall g, \forall t
\end{eqnarray}
\begin{eqnarray}
-\pi \leq \theta_{i, t} \leq \pi \hspace{1cm} \forall i, \forall t
\end{eqnarray}
\begin{eqnarray}
|P_{i, t}| \leq P_{l}^{max}  \hspace{1cm} \forall i, \forall t
\end{eqnarray} 
\begin{eqnarray}
P_{i, t} = -B_{ij}(\theta_{i} - \theta_{j}) \hspace{1cm} \forall l, \forall t, \forall i, \forall j
\end{eqnarray} 
\begin{eqnarray}
p_{i, t} = \sum_{j=1}^{L_{i}} B_{ij}(\theta_{i} - \theta_{j}) \hspace{1cm} \forall i, \forall t
\end{eqnarray} 
\begin{eqnarray}
P_{i, t} = \sum_{g \in G_{i}}^{}P_{g, t} - P_{d, i, t} + P_{c, i, t}\hspace{1cm}\forall g, \forall i, \forall t
\end{eqnarray} 

The objective function (7) represents the total operating cost over a 24-hour period and consists of the generator operating cost and the load shedding cost.
 The generator operating cost is calculated as the product of the cost of generating power and the amount of power generated during that time.
  Operating costs are shown as a second-order function of generation. The load shedding cost is the cost of not meeting demand, 
  which is set very high to minimize the occurrence of load cuts. 
Equation (8) limits the generator's output to within its minimum and maximum generation capacity. A constraint on the phase angle difference to regulate power flow
 between points is imposed as (9). 
Equation (10) is a line power constraint, which limits the maximum amount of power that can travel over a line.
 The calculation includes a cutoff value because power can flow in both directions. The value of $P_{l}^{max}$ is constant over time for the SLR method,
  but changes in real-time to reflect weather conditions for the DLR method. 
Equation (11) calculates the power flow through the line and is composed of the product of the admittance matrix elements and the phase angle difference.
 Equations (12) and (13) represent the power flow equations to ensure that the supply and demand of power must be balanced at each hour.

\subsection{Simulation Method}
The main goal of this study is to find a trade-off between the safety and efficiency limitations of the existing weather criteria and the complexity of DLR facilities.
 For this purpose, the simulation was performed by updating the weather conditions according to the following criteria.
 \begin{table}[tb]
\begin{center}
\caption{Criteria for updating weather conditions}
\label{tab:Table1}
\begin{tabular}{c|p{6.8cm}}
\hline
\textbf{Symbol} & \textbf{Criteria} \\
\hline
$SLR_{Conv}$ & Conventional weather criteria (based on 40℃ temperature and 0.5m/s wind speed) \\
\hline
$SLR_{1month}$ & Weather criteria updated monthly\\
\hline
$SLR_{Season}$ & Updated weather criteria for each season (December-February: Winter, March-April/October-November: Spring/Autumn, May-September: Summer) \cite{sungduk}.\\
\hline
$SLR_{6month}$ & Updated weather criteria every 6 months (November through April, May through October) \cite{sungduk}. \\
\hline
$DLR_{10\%}$ & Hourly updated weather criteria are applied to 10\% of all transmission lines. \\
\hline
\end{tabular}
\end{center}
\end{table}

This study conducted a simulation of a 30-bus system \cite{PAS} and solved the optimization problem using Python's Pyomo library and Cplex solver.
In the simulation, hourly temperature data (8760 hours in total) of the Miryang area in 2023 were used,
the wind speed was consistently set to 0.5m/s \cite{Data}, and the load shedding cost was set to \$9,000/MWh \cite{gen_cost}. 
The DC power flow is assumed, and the minimum generation capacity is set to 0 for calculation convenience.

Based on a normal distribution, there is a 0.3\% probability of reaching the maximum temperature, which is conservative enough\cite{normal}.

 In this study, the concept of demand ratio was introduced to change the demand every hour, and the load data of each bus 
 was multiplied by the demand ratio to change the demand of all buses.
To account for changes in transmission capacity similar to changes in demand, we adopted a method of applying a transmission capacity ratio 
to each hour. Specifically, the maximum transmission capacity of each line is multiplied by the transmission capacity ratio to reflect 
the change in transmission capacity when weather criteria changes.
In addition, when selectively applying DLR to certain lines, it is assumed that DLR is applied to 10\% of all lines.
The study's updated criteria for new weather conditions was selected by analyzing monthly maximum temperature data from 1974 to 2023. 
Using the historical data, each month's standard deviation was calculated, and then three times the standard deviation 
was added to the maximum monthly temperature to make it more conservative. Table 5. shows the temperature criteria for 
$SLR_{conv}$, $SLR_{1month}$, $SLR_{season}$, $SLR_{6month}$, and $DLR_{10\%}$ proposed in this paper.

\begin{table}[tb]
\begin{center}
\caption{Air Temperature Criteria (°C) for $SLR_{conv}$, $SLR_{1month}$, $SLR_{season}$, $SLR_{6month}$, and $DLR_{10\%}$}
\label{tab:Table3}
\begin{tabular}{c|c|c|c|c|c}
\hline
\textbf{Month} & \textbf{SLR\textsubscript{Conv}} & \textbf{SLR\textsubscript{1month}} & \textbf{SLR\textsubscript{Season}} & \textbf{SLR\textsubscript{6month}} & \textbf{DLR\textsubscript{10\%}} \\
\hline
1 & 40 & 24.37 & 32 & 36.95 & Changes hourly\\
\hline
2 & 40 & 32 & 32 & 36.95 & Changes hourly\\
\hline
3 & 40 & 32.61 & 36.95 & 36.95 & Changes hourly\\
\hline
4 & 40 & 36.95 & 36.95 & 36.95 & Changes hourly\\
\hline
5 & 40 & 42.91 & 45.11 & 45.11 & Changes hourly\\
\hline
6 & 40 & 40.25 & 45.11 & 45.11 & Changes hourly\\
\hline
7 & 40 & 45.11 & 45.11 & 45.11 & Changes hourly\\
\hline
8 & 40 & 44.65 & 45.11 & 45.11 & Changes hourly \\
\hline
9 & 40 & 41 & 45.11 & 45.11 & Changes hourly\\
\hline
10 & 40 & 34.87 & 36.95 & 45.11 & Changes hourly\\
\hline
11 & 40 & 32.86 & 36.95 & 36.95 & Changes hourly\\
\hline
12 & 40 & 26.22 & 32 & 36.95 & Changes hourly\\
\hline
\end{tabular}
\end{center}
\end{table}

\subsection{Simulation Results}
Table 6. shows the monthly power system operating costs as a result of the simulation.
\begin{table}[tb]
\begin{center}
\caption{Operating costs for Rating Type(per \$1,000)}
\label{tab:Table4}
\begin{tabular}{c|c|c|c|c|c}
\hline
\textbf{Month} & \textbf{SLR\textsubscript{Conv}} & \textbf{SLR\textsubscript{1month}} & \textbf{SLR\textsubscript{Season}} & \textbf{SLR\textsubscript{6month}} & \textbf{DLR\textsubscript{10\%}} \\
\hline
1 & 444,585 & 387,180 & 412,939 & 432,002 & 442,720\\
\hline
2 & 330,208 & 303,201 & 303,646 & 319,441 & 328,598\\
\hline
3 & 300,762 & 273,490 & 288,815 & 288,815 & 298,381\\
\hline
4 & 246,634 & 236,357 & 236,674 & 236,674 & 244,292\\
\hline
5 & 245,231 & 255,944 & 265,138 & 265,138 & 243,336\\
\hline
6 & 295,147 & 296,788 & 315,620 & 315,620 & 293,293\\
\hline
7 & 456,293 & 478,574 & 478,643 & 478,643 & 454,555\\
\hline
8 & 407,689 & 427,266 & 429,880 & 429,880 & 405,919\\
\hline
9 & 297,843 & 300,809 & 316,811 & 317,945 & 296,164\\
\hline
10 & 282,547 & 263,189 & 270,913 & 302,847 & 280,230\\
\hline
11 & 325,947 & 298,861 & 313,662 & 314,277 & 323,917\\
\hline
12 & 446,468 & 394,097 & 414,773 & 433,853 & 444,737\\
\hline
\end{tabular}
\end{center}
\end{table}

Figure 5. shows a relative comparison of the operating costs of the $SLR_{conv}$ method with the other criteria set to 100.

\begin{figure}[tb]
\begin{center}
\includegraphics[width=\linewidth]{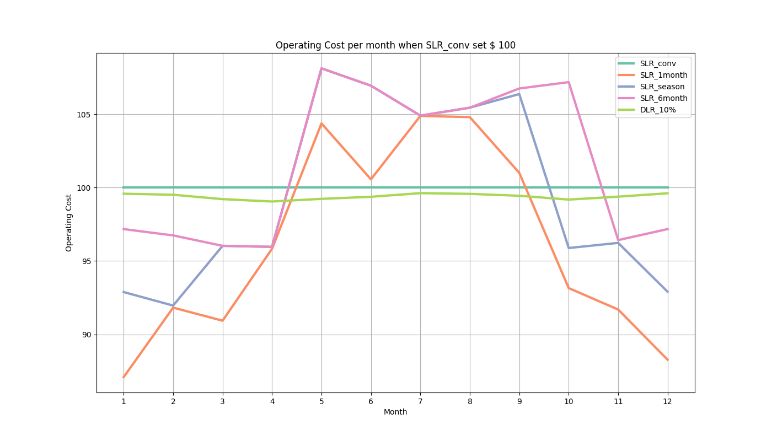}
\end{center}
\caption{Monthly operating cost comparison with SLR cost set to 100}
\label{fig:Figure5}
\end{figure}

As Figure 5. shows, in most of the months, the grid can be operated at a lower cost compared to $SLR_{conv}$ method. 
However, in some months, operating cost increases compared to the $SLR_{conv}$ method, which is due to the more conservative weather modeling and
reduced line capacity.
Based on this, the operating costs of the grid were compared over the course of a year. 
Figure 6. shows the power system operating costs over a year for each method. 

\begin{figure}[tb]
\begin{center}
\includegraphics[width=\linewidth]{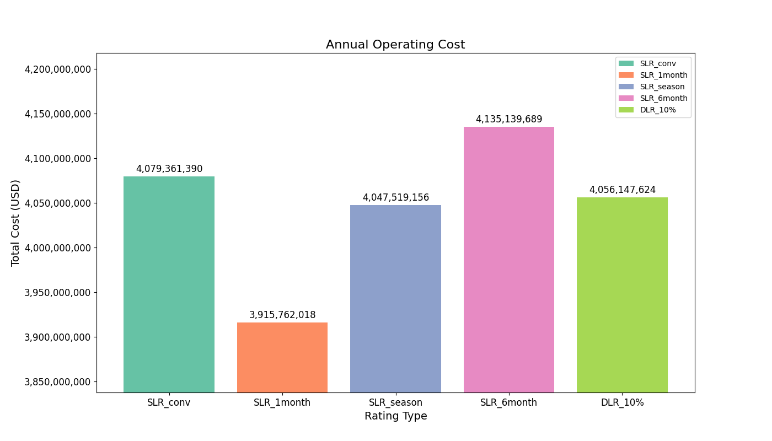}
\end{center}
\caption{Operating cost analysis in each case(by year)}
\label{fig:Figure6}
\end{figure}

From figure 6., it can be seen that $SLR_{6month}$ method has higher operating costs than $SLR_{conv}$ method, but $SLR_{1month}$ method,
$SLR_{season}$ method, and $DLR_{10\%}$ have lower operating costs than $SLR_{conv}$ method. 
This is due to the temporary increase in operational costs due to more conservative weather conditions when temperatures are high.
However, it suggests that the overall operating cost could be reduced by relaxing the weather conditions at other times. 
This study proposes $SLR_{1month}$ method, which updates weather criteria on a monthly basis, which can improve reliability 
by applying more conservative weather criteria during warmer months, and improve efficiency by relaxing weather criteria during cooler months. 
It is more economical to apply $SLR_{1month}$ method than to selectively operate DLR on certain lines, 
as it is less costly to operate the power system. $SLR_{1month}$ method overcome the technical limitations of $DLR_{10\%}$ 
and compensate for the limitations of conventional SLR weather criteria, providing an effective compromise that balances safety and economics.

\section{Conclusion}
This paper analyzes various limitations of the conventional SLR weather criteria operating in South Korea. 
Various weather updating approaches were explored in the study to find a balance between the conventional limitations of SLR and those of DLR:
$SLR_{1month}$ with monthly updates, $SLR_{season}$ with seasonal updates, and $SLR_{6month}$ with six-monthly updates. 
The concept of OPF was applied to simulate the operating costs in each case. $SLR_{season}$ showed that the power system can be operated 
more efficiently and reliably compared to the conventional SLR, but the effect is small, which limits its practical application. 
Therefore, this study proposes that the SLR1month method, i.e., updating weather criteria on a monthly basis, is an optimal compromise 
that can overcome both the limitations of DLR and the limitations of the conventional SLR method.
However, this study is limited in that it does not consider the on/off state of the generator, 
the cost of the startup state and the minimum and maximum startup time of the generator. These factors should be considered in future studies.
In this study, the simulation was performed on a test system, but it can be replaced with a real system to increase the realism. 
In addition, Reliability analysis for transmission lines can be conducted during summer months.

\acknowledgment
This work was supported by the National Research Foundation of Korea(NRF) grant funded by the Korea government(MSIT)(RS-2023-00248054)


\begin{thebibliography}{99}
\bibitem{PowerFlowEq}
"The Power Flow Equations," detailed documentation on power flow equations emphasizing the admittance matrix and its significance in network analysis
\end{thebibliography}
\end{document}